# Engineering decomposition-resilience and de-coalescence of GaN nanowire ensembles

Swagata Bhunia[1,2], Ritam Sarkar[2], Dhiman Nag[2], Kankat Ghosh[2], Krista R Khiangte[1], Suddhasatta Mahapatra*[,1] and Apurba Laha*[,2]

[1]Department of Physics, Indian Institute of Technology Bombay, Mumbai-400076, India
[2]Department of Electrical Engineering, Indian Institute of Technology Bombay, Mumbai-400076, India
 **S:** *Supporting Information*

**ABSTRACT:** The rapidly increasing interest in nanowires (NWs) of GaN and associated III-Nitrides for (opto-)electronic applications demands immediate addressal of the technological challenges associated with both NW-growth and device processing. Towards this end, we demonstrate in this work an approach to suppress thermal decomposition of GaN NWs, which also serves to remedy the effect of NW-coalescence during growth. While both these effects are well-known to be major hurdles in the development of GaN-NW-devices, reliable methods to tackle these issues have not been reported so far. We show that by providing a thin AlN cap layer, which epitaxially grows only on the top-facet of the GaN NWs, thermal decomposition can be almost completely suppressed. Thermal annealing of GaN NW-ensembles, post AlN-capping, leads to selective decomposition of uncapped/partially-capped NWs, leaving behind (mostly) AlN-capped GaN NWs, with superior crystal- and luminescence characteristics. This simple yet extremely effective approach may therefore serve as a very crucial milestone in the roadmap of GaN-NW-based (opto-)electronic technology.

Gallium Nitride (GaN) nanowires (NWs) are rapidly emerging as promising candidates for a variety of optoelectronic applications, including single photon emission at room temperature.[1,2,3] As a semiconductor, GaN offers attractive material properties, such as, excellent thermal- and chemical-stability,[4] high dielectric-breakdown voltage,[5] the potential for band-gap-tuning over the entire solar-spectral range (by alloying with Indium and Aluminum),[6,7] large exciton-binding-energy,[8] and sufficiently-large band-offsets for room temperature exciton confinement (in alloy heterostructures).[9] The NW-geometry combines further advantages due to the high surface-to-volume ratio and the possibility of inhibiting propagation of threading dislocations.[10,11] Furthermore, the free surface of the NW-sidewalls allows the lattice strain to be released elastically.[12] All these factors are expected to improve light-matter-coupling and thus enhance photon absorption and extraction efficiency of NW-based photonic elements. Thus, GaN and related alloy NWs are being extensively explored in the context of fabricating light emitting diodes (LEDs),[13] ultra-low-threshold polariton lasers,[14] solar cells,[15] and for photo-electrochemical water splitting.[16] By embedding single quantum dots in III-Nitride NWs, single photon emission at room temperature has also been demonstrated, very recently.[3]

While the attractive material and optical characteristics of GaN NWs are well-recognized now, several technological challenges remain to be addressed, towards the goal of realizing practical NW-based optoelectronic devices. For example, stabilizing GaN NWs against decomposition at high temperatures and suppressing their coalescence (in dense ensembles) are important capabilities to develop, to explore their full application-potential. For device processing steps such as dopant activation and formation of low resistance Ohmic contacts, high temperature annealing of GaN (and other III-Nitride heterostructures) is very crucial.[17,18] In the context of Mg-doping of GaN epilayers, annealing at 850°C ensures optimal breaking of the Mg-H bonds in metal organic chemical vapor deposition (MOCVD) grown layers, thus enabling Mg incorporation.[19] For molecular beam epitaxy (MBE) growth, annealing at 600°C induces dissociation of the Mg-O bonds and assists in increasing the doping concentration.[20] Moreover, contacts to GaN made with Ti/Al/Ni/Au metal stacks exhibit sufficiently Ohmic nature, only when they are annealed at 800°C and above.[18] Finally, earlier reports have demonstrated that annealing at temperatures between 1227°C and 1427°C (under $N_2$ overpressure of 10 kbar) allows annihilation of Ga-vacancies,[21] which act as non-radiative recombination centers suppressing the luminescence efficiency of GaN NWs. However, at these elevated temperatures, GaN NWs start to decompose from their top-facets, thereby shrinking significantly in both length and diameter.[22]

Another challenge with both MOCVD and MBE growth of GaN NWs is the tendency of adjacent NWs to coalesce, within dense (~ $10^9$ – $10^{10}$ cm$^{-2}$) ensembles.[23,24,25,26] Coalescence in the initial stages occurs due to the radial growth of the NWs, while in later stages, it may be attributed to their out-of-plane orientational spread. Excess surface energy associated with the orientational spread, compared to the elastic energy of bending, promotes coalescence beyond a critical NW-length. In Ref. 26, coalescence of GaN NWs separated by ~ 60 nm was observed to take place beyond a length of ~340 nm. Coalescence not only increases the effective cross-sectional area of the NWs, but also creates a network of boundary dislocations[27,28] which, in turn, introduces non-radiative recombination centers. While elegant methods to estimate the degree-of-coalescence have been proposed,[29] techniques to inhibit or remedy the effect of NW-coalescence are yet to be developed.

In this Letter, we demonstrate that thermal decomposition of GaN NWs can be completely suppressed, by encapsulating the as-grown NWs with a thin AlN cap layer, prior to annealing. Our results reveal that AlN grows (epitaxially) only on the top-facet of the NWs, and yet it prevents any reduction in their length or diameter. In earlier reports, thermal decomposition was modeled considering the desorption from top-facets of the NWs, in a quasi-layer-by-layer mode.[22] Suppression of the decomposition process by AlN-capping not only validates this model but also establishes a robust, yet simple, technique for enhancing the thermal stability of GaN NWs significantly. Further, we demonstrate a remedial effect of the process of AlN-capping and subsequent thermal annealing, wherein the degree-of-coalescence is observed to reduce, with concomitant reduction of the NW fill-factor within the ensemble. These results thus represent crucial technology-enabling capabilities, towards realization of GaN-NW-based practical (opto-)electronic devices.

All samples (labeled A, B, C, D, E) were grown in a RIBER (Compact 21) plasma assisted molecular beam epitaxy (PAMBE) system, equipped with effusion cells for Ga and Al solid sources and a RIBER ADDON radio frequency (rf) plasma source for active Nitrogen. Prior to PAMBE growth, the sapphire (0001) substrates were chemically cleaned with trichloroethylene, acetone, isopropyl alcohol and DI water, in an ultrasonic bath. The substrates were then loaded into the load-lock chamber of the MBE system and heated at 100°C for 1 hr, with an infra-red lamp. Subsequently, the substrates were degassed overnight in the buffer chamber at 400°C and transferred to the growth chamber. Within the growth chamber, the substrate temperature ($T_S$) was first raised to 885°C to anneal the substrates for 1 hr at a background pressure of $3 \times 10^{-9}$ Torr. For nitridation of the substrates, $T_s$ was then reduced to 400°C and the plasma shutter was opened for 1 hr 20 min. During this nitridation step, the rf forward power and the N2 flow rate were maintained at 350 Watt and 2.4 sccm, respectively. With the plasma shutter open, $T_s$ was subsequently raised to 800°C (at a rate of 20°C/min) and thin (~6 nm) AlN layers were grown with an Al flux of $6.57 \times 10^{-8}$ Torr. For the GaN-NW-growth, $T_s$ was next raised to 880°C and the Ga and the nitrogen-plasma shutters were kept open for 3 hrs 30 min. This resulted in the growth of GaN NWs ~950 nm long. For sample E, another ~15 nm AlN was grown on the crest of the NWs at the same $T_s$. While for the GaN NW growth, the rf power and the N2 flow rate were maintained at 475 Watt and 3.5 sccm, respectively, for AlN layer growth the same parameters were kept at 240 Watt and 0.65 sccm, respectively.

For the annealing studies, all samples, except sample A, were annealed at $T_s$ = 950°C for different durations (at a background pressure of $5 \times 10^{-8}$ Torr). While samples B, C, and D (uncapped samples) were annealed for $t_A$ = 30 min, 1 hr, and 2 hr, respectively, the AlN-capped sample (sample E) was annealed only for 2 hours. The growth was monitored real-time by Reflection High Energy Electron Diffraction (RHEED). Microstructural properties of all the samples were characterized by Field Emission Scanning Electron Microscopy (FE-SEM) and High-Resolution Transmission Electron Microscopy (HR-TEM), and the optical properties were studied by temperature-dependent photoluminescence (PL) measurements, carried out using a 325 nm-He-Cd laser of spot-size of diameter 80 μm.

Figure 1(a)-1(f) show RHEED images recorded along the $<1\bar{1}00>$ azimuth of sapphire, at different stages of the nitridation and NW-growth processes. The streaky pattern shown in Figure 1(a) indicates the formation of a smooth growth-front, after annealing of the substrate at 885°C. Figure 1(b) shows the streaky RHEED pattern obtained at the end of the nitridation process. This streaky pattern is characterized by a smaller inter-streak separation compared to that in Figure 1(a), and is contributed by the $<11\bar{2}0>$ azimuth of the AlN layer (since $[11\bar{2}0]_{AlN}||[1\bar{1}00]_{Al_2O_3}$[30]). Absence of any oxide-related streaks in the image confirms complete nitridation of the surface. Figure 1(c) shows a streaks-with-spots RHEED pattern obtained after growth of 6 nm AlN at $T_s$=800°C. The pattern along the same azimuth became completely spotty (not shown here), within barely one minute of GaN growth at $T_s$=880°C, indicating the nucleation of GaN NWs. This spotty pattern persisted (but with evolving spot size and shape) till the end of GaN-growth (total duration of 3 hr 30 minutes), as revealed by Figure 1(d).

A comparison of the RHEED images of Figures 1(e) and 1(f) provides the first evidence of the fact that AlN-capping of the GaN NWs indeed inhibits their decomposition of the latter, during subsequent thermal annealing. Figure 1(e) shows the spotty pattern corresponding to the uncapped NWs, after annealing for 2hr at 950°C. It is evident that the spots of this image are broader than those of Figure 1(d), strongly suggesting annealing-induced thinning of the NWs. On the other hand, the RHEED pattern of Figure 1(f), recorded after annealing of the AlN-capped GaN NWs under the same conditions, shows negligible change in the size of the RHEED spots, in comparison those in Figure 1(d). This indicates that the diameter of the GaN NWs did not change significantly. The blurriness and apparent broadening of the RHEED spots in Figure 1(f) may be attributed to the presence of the AlN cap layer, as discussed in more detail in the context of the inset in Figure 4(a).

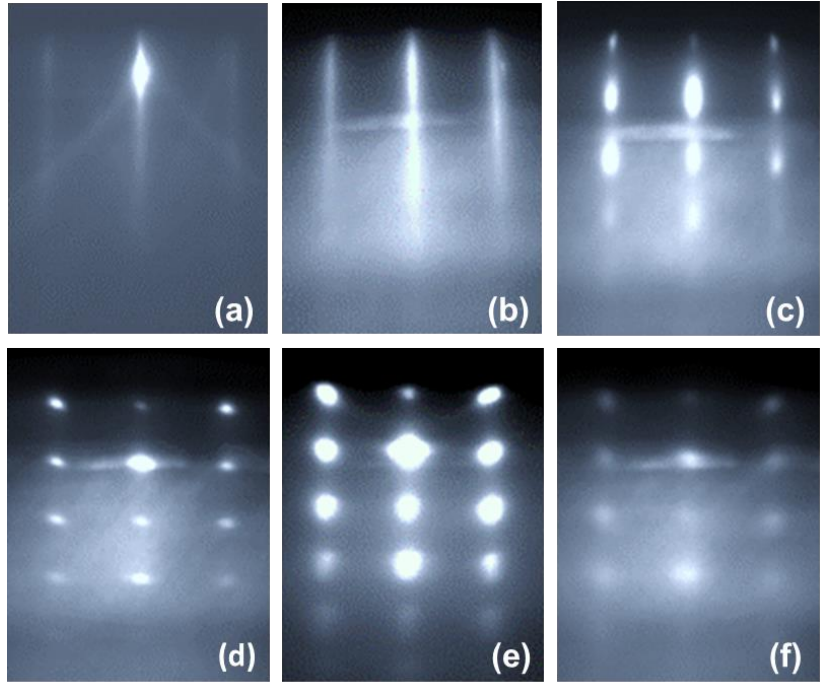

**Figure 1.** RHEED patterns recorded (a) along the $<1\bar{1}00>$ azimuth of Sapphire, before nitridation, (b) along the $<11\bar{2}0>$ azimuth of AlN, recorded at the end of the nitridation process and (c) along the same azimuth of AlN, after 8 min of AlN growth. The spotty RHEED pattern along the $<11\bar{2}0>$ azimuth of GaN (d) after 3 hrs. 30 min of GaN growth, (e) after annealing at 950°C for 2 hrs. of the uncapped NWs, and (f) after annealing at 950°C for 2 hrs. of the AlN-capped NWs at a background pressure of 5x10-8 Torr.

SEM imaging of the NWs (Figure 2) provides a direct probe of the decomposition phenomenon. Figures 2(a), 2(b), and 2(c) show the 45°-tilted-view images of the NWs corresponding to samples A (uncapped/un-annealed), D (uncapped, annealed) and E (AlN-capped, annealed), respectively, while Figures 2(d), 2(e) and 2(f) show the corresponding top-view images. Annealing-induced reduction in length and diameter of the NWs is clearly observed, by comparing the SEM images corresponding to samples A and D. On the other hand, no significant change of the NW-dimensions due to annealing is observed in case of sample E. The average length (diameter) of NWs in samples A, D, and E were estimated to be 953 nm (68 nm), 265 nm (10 nm), and 900 nm (66 nm), respectively. Evidently, the resilience of the GaN NWs against thermal decomposition is strongly enhanced due to AlN-capping.

To quantitatively probe the temporal evolution of NW-decomposition, we carried out annealing of the uncapped GaN NWs at $T_s$ = 950°C, for different durations. According to the model developed by Zettler et al.,[22] the decomposition rate of a GaN NW ($\Phi_{des}$) may be expressed as

$$\Phi_{des} = -C_2 \frac{\partial V}{\partial t}$$

where, $V$ is the volume of an individual NW, and $C_2$ is a characteristic constant. In this model, the decomposition process is assumed to take place in a layer-by-layer mode, as illustrated in Figure 2(g). The decomposition of the NWs commences at the intersection of the side-wall edges and the top-facet, creating kinks which propagate along the periphery of the latter. Eventually the top-facet is completely decomposed, before kinks start to appear in the layer below.[31] Furthermore, a cylindrical NW geometry is considered for analytical simplicity. The time-evolution of the radius ($r$) and the height ($h$) of the NWs is then expressed by

$$\frac{\partial r}{\partial t} = -R_s \pi r \text{ and } \frac{\partial h}{\partial t} = -2R_t \pi r$$

where, $R_s$ and $R_t$ are the rate-constants for decomposition of the sidewalls and the top-facet of the NW, respectively. From these equations we obtain $r(t) = r_0 exp(-R_s \pi t)$ and $h(t) = h_0 - \frac{2R_t r_0}{R_s}(1 - exp(-R_s \pi t))$. Figure 2(h) shows the experimentally-determined dependence of the average values of $r$ and $h$ on the annealing time. These values were estimated by analyzing the corresponding SEM micrographs with the image processing software ImageJ.[32] The obtained histograms for $r$ and $h$ are shown in Supplementary Figure S1. Fits of the experimental data with the obtained expressions for $r(t)$ and $h(t)$ are depicted by the solid lines in Fig. 2(h). From these fits, we determined the decomposition rate constants as $R_s = 7.2 \times 10^{-5}$ sec$^{-1}$ and $R_t = 7.6 \times 10^{-4}$ sec$^{-1}$. These values are approximately 30% of the corresponding rate constants estimated by Zettler et al.,[22] by direct measurement of the Ga flux during annealing (at 920°C), using a quadruple mass spectrometer. As reported in Ref. 22, these GaN NWs were grown directly on Si(111) substrates. It may therefore be inferred that GaN NWs grown on (AlN-nitridated) sapphire substrates are less prone to annealing-induced decomposition. Figure 2(h) also shows the average radius and height of the AlN-capped GaN NWs (violet square and violet sphere, respectively), annealed for 2 hr at 950°C (sample E). Clearly, no significant change in either of the parameters can be seen, when compared to those of the as-grown NWs (sample A). Thus, the SEM analysis convincingly establishes the fact that AlN- capping stabilizes GaN NWs against annealing-induced decomposition.

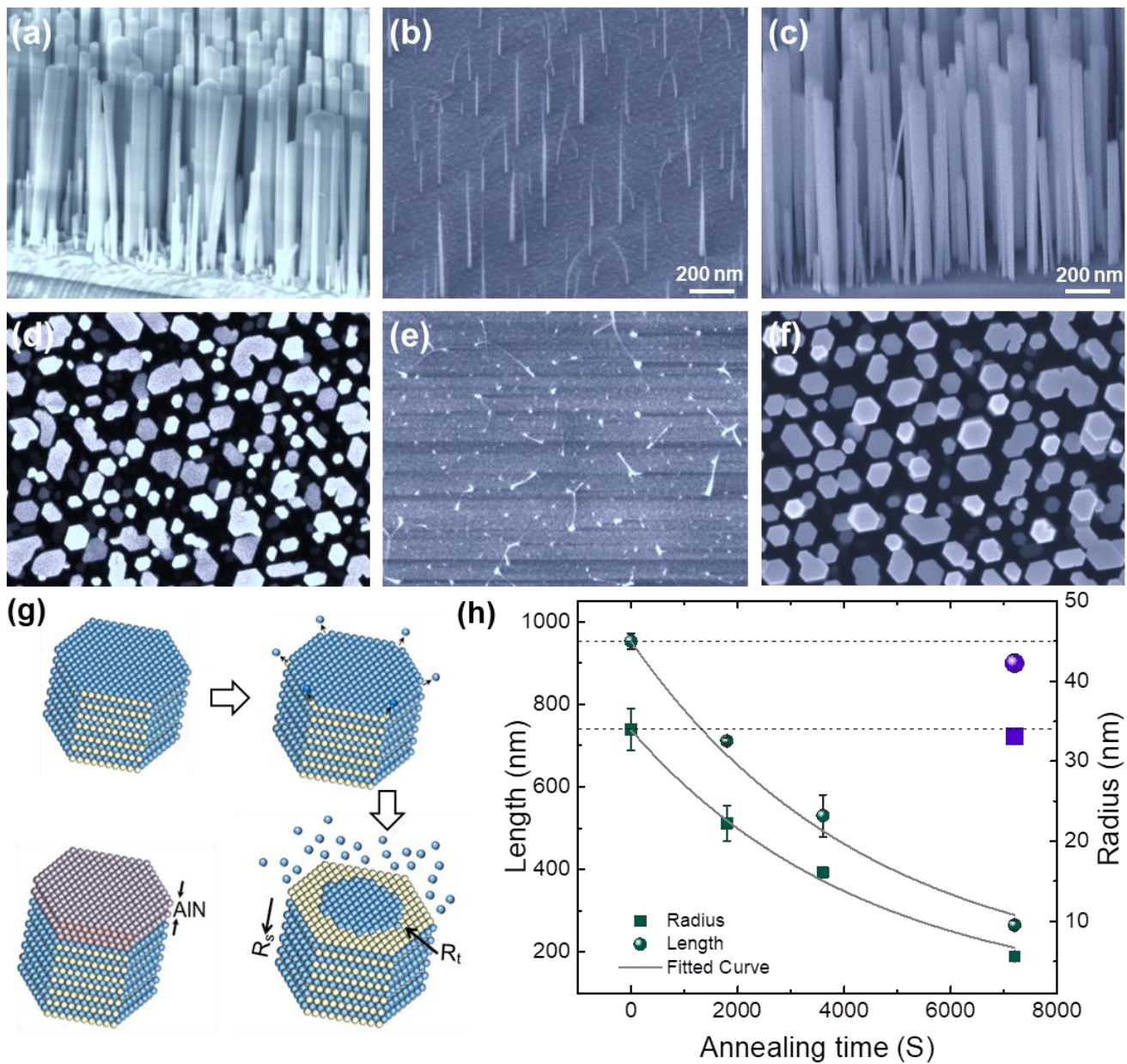

**Figure 2.** SEM (a)-(c) 45°tilted (bird view) view and (d)-(f) top-view of the NWs, corresponding to samples A, D, and E, respectively. These images were processed with ImageJ software to obtain length and diameter distributions, as well as their circularity and perimeter distributions. (g) Schematic representation of the NW-decomposition process (clockwise from top-left), and depiction of decomposition-inhibition due to AlN-capping of GaN NWs (bottom-left). (h) Plot showing the change in the average length (teal spheres) and radius (teal squares), with respect to the duration of annealing. The solid lines represent the fitting of the data with the equations; $h = h_0 - \frac{2R_t r_0}{R_s}(1 - exp(-R_s \pi t))$ and $r = r_0 exp(-R_s \pi t)$. Also shown in the figure are the average length (violet sphere) and radius (violet square) of the AlN-capped GaN NW-ensemble, after annealing at 950°C for 2 hrs.

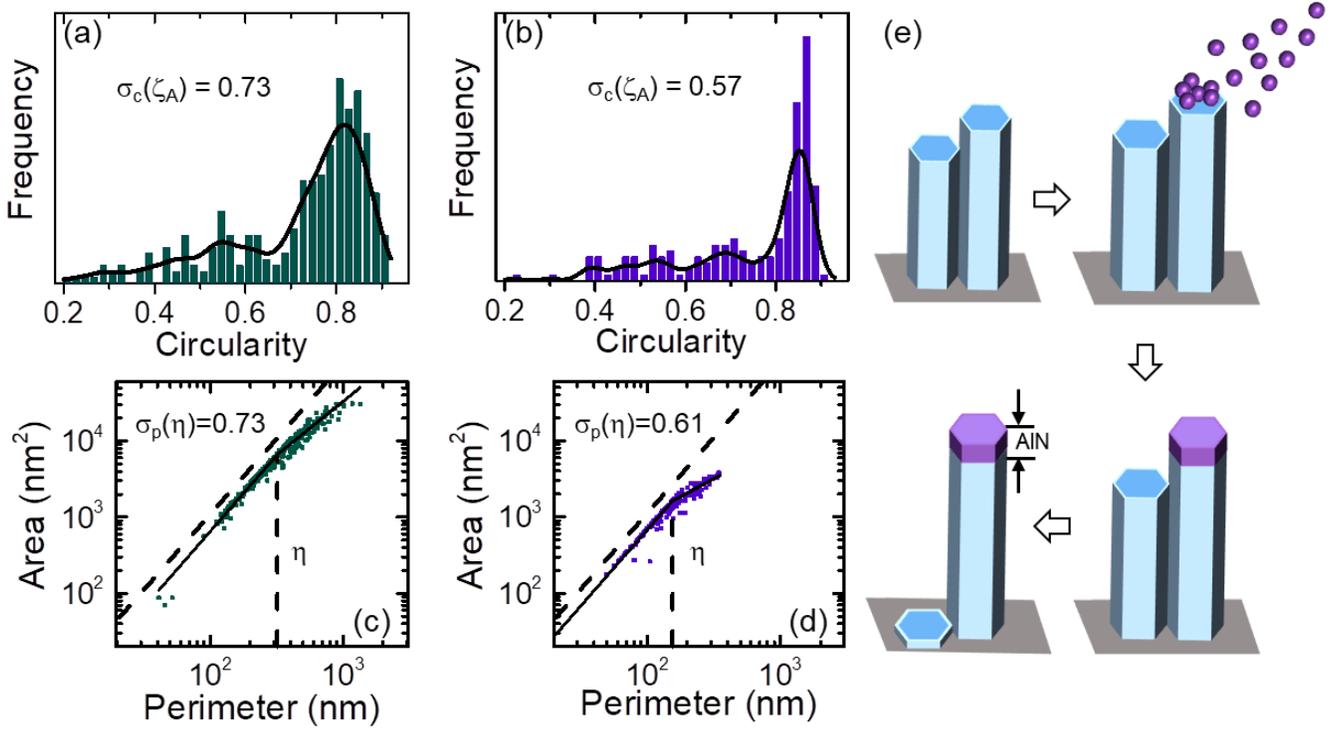

**Figure 3.** Circularity histograms of NW-ensembles, corresponding to (a) sample A and (b) sample E. The variation of the top-facet area with their perimeter, for NW-ensembles of (c) sample A and (d) sample E. (e) Schematic illustration depicting the "de-coalescence" effect due to shadowing of the molecular beam by the taller NWs (top-right), preventing AlN capping of the shorter ones (bottom right). The latter decomposes during subsequent annealing (bottom left).

Besides providing thermal stability, the process of AlN-capping and subsequent annealing yields ensembles of GaN NWs with reduced degree-of-coalescence compare to their as-grown and uncapped counterparts. Brandt et al.[29] have recently developed an algorithmic procedure to estimate the degree-of-coalescence in GaN NW-ensembles, which is based on the analysis of the area and perimeter of the NW top-facets, as observed in plan-view SEM images (similar to Figures 2(d) and 2(f)). In this model, the degree-of-coalescence ($\sigma$) is defined in two different ways: (a) the fractional top-facet cross-sectional area, $(\sum A_{C<\zeta})/A_T$, with circularity (C) below a threshold value ($\zeta$), and (b) the fractional cross-sectional area, $(\sum A_{P>\eta})/A_T$, of top-facets with perimeter (P) above a critical value ($\eta$), termed as $\sigma_C$ and $\sigma_P$, respectively. Here, circularity of an individual NW is defined as $C = 4\pi A/P^2$, and the threshold circularity $\eta$ refers to the value of C calculated for the top-facet obtained by coalescence of two regular-hexagonal NWs, with the coalescence-boundary parallel to either the A-plane ($\eta_A = 0.762$) or the M-plane ($\eta_M = 0.653$). Circularity histograms of NWs corresponding to sample A (uncapped and unannealed) and sample E (AlN-capped and annealed), along with plots of kernel density estimations (black solid lines) are shown in Figures 3(a) and 3(b), respectively. While in both cases, C is peaked at values corresponding to single NWs, the relative frequency of low-circularity NWs is lower for sample E. Correspondingly, the estimated degree-of-coalescence is observed to drop from $\sigma_C(\eta_A) = 0.73$ for sample A to $\sigma_C(\eta_A) = 0.57$ for sample E.

Figures 3(c) and 3(d) depict the alternate area-perimeter representation, wherein the perimeter-dependence of the top-facet area changes abruptly from a quadratic to a linear behavior, at the critical perimeter $\eta$ (represented by the dotted lines). The critical perimeter is obtained from a fit of the data with a Heaviside scaling function, as described in Ref. 29. The estimated degrees-of-coalescence ($\sigma_P$ in this case) from these plots are very close to the values of $\sigma_C(\eta_A)$ obtained above. More notably, the coalescence degree of sample E ($\sigma_C(\eta_A) = 0.57, \sigma_P = 0.61$) is lower than the lowest reported in Ref. 29 ($\sigma_C(\eta_A) = \sigma_P = 0.74$ and comparable to sample A), wherein coalescence of GaN-NW ensembles (on Si substrates) with different fill-factors was studied. This suggests that the process of AlN-capping and subsequent annealing allows reducing $\sigma$ significantly below the values typically obtained in as-grown and uncapped GaN NW ensembles.

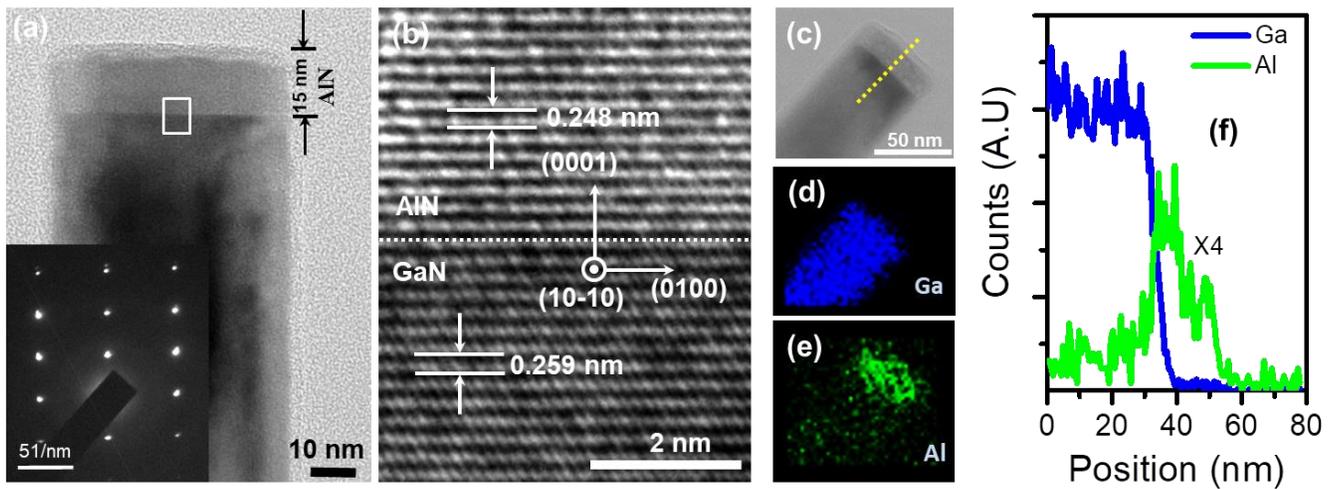

**Figure 4.** (a)The bright field TEM image of the AlN capped GaN NW, revealing epitaxial growth of AlN on GaN along the (0001) direction. The inset shows the SAED pattern, wherein two super-imposing diffraction patterns of the same symmetry can be seen, due to presence of both GaN and AlN in the sampled region. (b) High-resolution TEM image showing the (0001) planes of the AlN and the GaN regions across the sharp AlN-GaN boundary. For the region shown in the image of (c), the recorded EDX-based composition maps for (d) Ga and (e) Al reveal nearly no-intermixing across the interface. (f) The composition line profiles along the yellow dotted line in (c).

Coalescence sets in even at early stages of NW-growth, while AlN-capping in sample E was performed at the end of the growth process. What leads to the reduction of σ in case of the AlN-capped GaN-NW-ensemble is therefore an intriguing point to investigate. To probe this, we first note that the NW-fill-factor in the ensemble of sample E ($3.14 \times 10^9 cm^{-2}$) is ~ 59% of that of sample A ($5.29 \times 10^9 cm^{-2}$).

This suggests that nearly half of the as-grown NWs within the ensemble of sample E underwent decomposition, as a result of thermal annealing. The reason behind this counter-intuitive phenomenon may be understood from the schematic depiction of the AlN capping process, shown in Figure 3(e). Due to the variation in their heights, not all NWs in the as-grown ensemble are uniformly capped by AlN. Particularly, NWs within coalesced aggregates (resulting from late-stage coalescence) and/or thicker NWs (resulting from early-stage coalescence) are likely to be shorter than individual-NWs with nearly-hexagonal top-facets. In such a scenario, the shadowing of the Al/N molecular beams by the taller individual-NWs hinders AlN-capping of the shorter ones. During subsequent annealing, these uncapped/partially-capped NWs decompose, leaving behind an ensemble of mostly AlN-capped NWs, with reduced fill-factor of the ensemble, but higher circularity of the individual NWs. This is indeed what is revealed by the histograms presented in Figures 3(a) and 3(b), together with the estimate of the fill-factors. The process of AlN-capping and subsequent annealing thus acts as a filter for retaining thermally stable NWs of higher circularity, which are also expected to be free from coalescence-induced dislocation networks.

Further understanding of how the decomposition process is inhibited due to AlN-capping is provided by HRTEM analysis (Figure 4). The TEM image of Figure 4(a) shows the 15 nm AlN cap layer, on top of the GaN NW. No signature of AlN growth along the sidewalls is visible in this image. The measured length (870 nm) and diameter (57 nm) of the NW match closely with the corresponding values obtained by analysis of the SEM images. The selected area electron diffraction (SAED) pattern, shown in the inset of Figure 4(a), depicts two nearly-superimposing sets of diffraction spots, with the same symmetry. Here, the larger spots are due to the GaN (0001) planes, while the smaller spots correspond to the (0001) planes of AlN. The recorded diffraction pattern suggests that the AlN cap layer is entirely single-crystalline and epitaxially related to underlying GaN NW. The close proximity of the spots due to AlN and GaN in the SAED image also explains the apparent broadening of the RHEED spots in Figure 1(f). Unlike in HRTEM, the diffracted electrons in RHEED are of low energy and consequently, the spots due to AlN and GaN regions are not as well-resolved. The bright field HRTEM image of Figure 4(b) reveals that AlN is (0001) oriented, with a sharp interface separating the two regions. The inter-planar spacings ($d$) between the (0001)-planes in the GaN and the AlN regions (0.259 nm and 0.248 nm, respectively) were measured to be equal to their bulk values. This strongly suggests that annealing at 950°C for 2 hr did not induce significant intermixing of the GaN and AlN regions.

Additionally, elemental analysis of the NWs around the AlN-cap region was performed by employing energy dispersive x-ray (EDX) point measurement in the TEM set-up. Figures 4(d) and 4(e) show the maps corresponding to elemental Ga and Al, recorded for the area shown in the TEM image of Figure 4(c). The line-profiles of Ga and Al contents, measured along the yellow dotted line in Figure 4(c) is shown in Figure 4(f). The maps and the line profile also reveal that (a) no significant inter-diffusion takes place at the NW-cap interface and (b) no observable overgrowth of the NW-sidewalls with AlN occurs, either during the capping process or during subsequent thermal annealing.

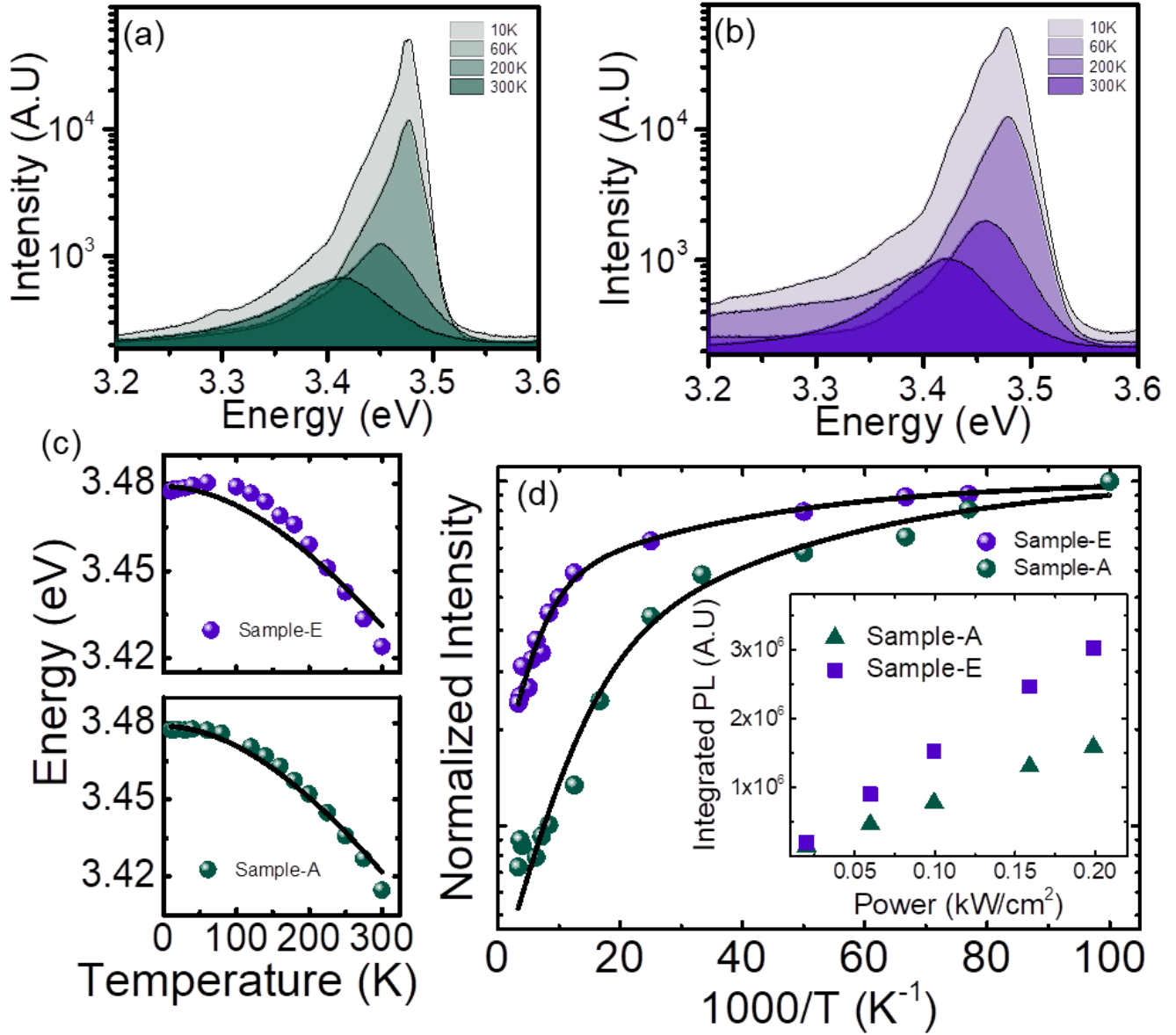

**Figure 5.** Photoluminescence (PL) spectra, recorded at a few temperatures, for the NW-ensembles of (a) sample A and (b) sample E. Corresponding variation of the near-band-edge (NBE) (c) emission energy and (d) emission intensity with temperature. The values in (c) were fitted with Varshni's equation (solid black line). Inset of (d) Excitation-power-dependence of the integrated-PL-intensity of sample A (uncapped, unannealed) and sample E (AlN-capped, annealed), performed at 10K.

In order to study the impact of thermal annealing on the optical properties of the NWs, temperature dependent photoluminescence (PL) measurements were carried out, for the NW-ensembles of samples A and E. Spectra recorded at a few temperatures for sample A and sample E are shown in Figures 5(a) and 5(b), respectively. The energy of the near-band-edge emission (from free exciton A, FX(A)) was obtained for each temperature by fitting the corresponding PL spectrum with Gaussian line shape functions (with 99% confidence bound) (See supplementary Figure S2). The temperature-dependence of the FX(A)-emission-energies are depicted in Figure 5(c), together with a fit of Varshni's equation,[33] $E(T) = E(0) - \alpha T^2/(\beta + T)$, where $E(T)$ is the transition energy at tem-perature $T$, and $\alpha$, $\beta$ are the Varshni's thermal coefficients. The best fit for sample A (sample E) was obtained for $E_A(0) = 3.479$ eV ($E_E(0) = 3.479$ eV), $\alpha_A = 7.05 \times 10^{-4}$ eV/K ($\alpha_E = 6 \times 10^{-4}$ eV/K), and $\beta_A = 814$K ($\beta_E = 828$K). These temperature variations of the FX(A) emission energies, reflecting the bandgap widening of bulk GaN, are comparable to values reported earlier.[34] The inset of Figure 5(d) shows the excitation-power-dependence of the integrated PL-intensity, estimated from the area under the PL response curve recorded at 10K, for both samples A and E. The observed linear-dependence implies that the PL emission is predominantly contributed by radiative recombination centers[35] and that exciton-exciton inelastic scattering processes may be ignored. For an excitation power of 0.2 kW/cm$^2$, the plots of near-band-edge emission intensities (obtained once again from the Gaussian fit of the corresponding PL-spectra) versus inverse-temperature are shown in the main panel of Figure 5(d). It is observed that the variation of the (normalized) integrated PL intensity with temperature is lower for the AlN-capped NWs, in comparison to that for the uncapped NWs, implying that the internal quantum efficiency (IQE) of the former ($I_{10K}/I_{300K}=0.23$) is greater than that ($I_{10K}/I_{300K}=0.08$) of the latter.

Enhancement of IQE with AlN-capping indicates to several beneficial aspects of our approach. Point defects are known to provide non-radiative decay channels for excitons.[36] Since the GaN NWs in this study were grown in N$_2$-rich conditions, Ga vacancies are expected to be the predominant type of point defects in these structures. As mentioned earlier, annealing of GaN NWs was shown to reduce the concentration of Ga-vacancy related defects.[21] Therefore, improvement of the IQE in sample E may also be attributed to annealing-induced annihilation of Ga-vacancies, enabled by AlN-capping of the NWs. Alternatively, /additionally, the higher IQE of sample E may also be explained by the "de-coalescence" phenomenon discussed in the previous section, whereby extended defects of coalesced-NW-aggregates are eradicated. Thus, by addressing two different issues with rather complementary origins, our approach of AlN capping and subsequent thermal annealing of GaN NWs bodes extremely well for improving the optical response of GaN NWs.

In conclusion, we demonstrated that by capping GaN NWs with a thin AlN layer, thermal decomposition of the former can be completely suppressed. This capability is expected to enable crucial device processing steps such as dopant incorporation and formation of low-resistance Ohmic contacts. Furthermore, by enabling in-situ post-growth annealing, this approach will also serve to enhance the optical response of GaN NWs.[22] Equally significant in this context is the "de-coalescence" effect, which eliminates channels for non-radiative recombination, by annihilation of dislocation networks.[29] Ultrathin NWs, comparable to the exciton-Bohr-radius (~3 nm) in GaN, would promote on demand single photon emission (in quantum-dot-in-NW structures[3]) by eliminating the bi-exciton state. Our approach is the first of its kind which holds the promise of stabilizing the GaN NWs, in this quantum regime.

## ASSOCIATED CONTENT

**S: Supporting Information**

The details calculation of NWs length and radius from the SEM images and the fitting of 10K PL spectra for sample A and sample E. This material is available free of charge via the Internet at http://pubs.acs.org.

## AUTHOR INFORMATION

**Corresponding Authors**
*E-mail: laha@ee.iitb.ac.in
*E-mail: suddho@phy.iitb.ac.in

**Author Contributions**
The manuscript was written through contributions of all authors.

## ACKNOWLEDGMENT

All authors acknowledge the financial support of the Ministry of Electronics and Information Technology (MeitY) for this work and the Centre of Excellence in Nanoelectronics (CEN), IITBNF facility for the technical support, towards the execution of this project. Swagata Bhunia acknowledges the financial assistant of the University of Grand Commission, The Govt. of India.